\documentstyle[12pt]{article}
\title{Cosmological Theories From $SO(2,2)/SO(2)\times SO(1,1)$\footnote
{DTP-MSU/97-03, hep-th/yymmnn}}

\author{L. A. Pando Zayas\thanks{e-mail address:leopoldo@grg1.phys.msu.su}\\
Department of Theoretical Physics,\\
Moscow State University,\\
Moscow 119899, Russia}

\begin{document}
\def\pp{\partial_{z}}
\def\ppm{\partial_{\bar{z}}}
\def\pd{\partial_{\mu}}
\def\pdn{\partial_{\nu}}
\def\pu{\partial^{\mu}}
\def\sh{\sinh}
\def\ch{\cosh}
\def\intt{\int\limits_{\Sigma}d^{2}z Tr}
\def\innt{\int\limits_{\Sigma}d^{2}z}
\def\tg{\tilde{g}}
\def\b{\beta}
\def\a{\alpha}
\def\Qp{Q_{+}}
\def\Qm{Q_{-}}
\def\S{\Sigma}
\def\T{1+t^{2}}
\def\te{\theta}
\def\D{\Delta}
\def\l{\Lambda}
\def\Ap{A_{z}}
\def\Am{A_{\bar{z}}}
\def\Jp{J_{z}}
\def\Jm{J_{\bar{z}}}
\def\Ip{I_{z}}
\def\Im{I_{\bar{z}}}
\maketitle

\begin{abstract}
We herein set forth intrinsically four-dimensional string
solutions and analyze some of its properties. The solutions are constructed
as gauged WZW models of the coset $SO(2,2)/SO(2)\times SO(1,1)$. We recover
backgrounds having metric and antisymmetric tensors, dilaton fields and two
electromagnetic fields.  The theories describe anisotropically expanding and
static universes for some time values.

\end{abstract}

\section{Introduction}
Recently constructing exact
string solutions on gravitational backgrounds \cite{tsey1} has interested
many string theorists, the main motivation of which has been that
these backgrounds can shed light on different areas of gravitation such as
the nature of singularities and black hole physics. Most of the found
solutions are constructed as the product of two theories.  One
element that is most frequently included in these constructions is
the two--dimensional, dilatonic black hole.  A remarkable step was taken by
Johnson who, by means of heterotic coset models \cite{hetco}, was able to
move away from the typical solution constituted by the two--dimensional
dilatonic black hole in the radial sector and a $SU(2)/U(1)$ theory in the
angular sector. Heterotic coset models provide a natural opportunity to
introduce twists between the radial and angular sectors.  The use of heterotic
coset models gives rise to the possibility of employing anomalously gauged WZW
models to construct a consistent theory. The
fact that there is a unique gauge extension of the WZW model having a
classical anomaly depending only on the gauge field \cite{witcom} is
fundamentally important in the construction of heterotic coset models.  This
anomaly cancelled against the chiral fermionic anomalies.  In this paper
we will use heterotic coset model techniques to construct solutions
that are intrinsically four--dimensional, meaning the coset is not the
product of lower dimensional cosets. This could potentially play a
fundamental role in studying the nature of singularities.  These
solutions reveal very interesting features from the gravitational
viewpoint.  Intrinsically four--dimensional solutions constructed from vector
gauged WZW models have been presented \cite{bars}, but without any
evident symmetry, and so their gravitational interpretation is not clear.
A complete list of cosets having a single time coordinate and up to ten
dimensions has been presented in \cite{gq}
based on equivalences between gauged WZW cosets and ordinary coset models.
In this paper we focus on one four--dimensional coset that has a single time
coordinate.  The classification presented  \cite{gq} does not give any
information on the structure of the background in which the string
propagates. The only relevant information gained from the
classification is the dimension of the space and that it has a single time
coordinate. Here we investigate the concrete structure of the background
corresponding to $SO(2,2)/SO(2)\times~ SO(1,1)$ and show that for some values
of the time coordinate this coset describes anisotropically expanding and
static universes.

\section{The $SO(2,2)/SO(2)\times SO(1,1)$ as a vector \break gauged WZW model}

The key construction of this section is the coset
$SO(2,2)/SO(2)\times SO(1,1)$ as it is defined by the diagonal vector gauging
of WZW  models. The action of the diagonal vector gauging of WZW model is a
particular case, corresponding to $A^{R}=A^{L}$, of the following the action:

\begin{eqnarray}
I(g,A^{R},A^{L})&=&kI(g)+kI_{A} \nonumber \\
I(g)&=&\frac{1}{4\pi}\intt \pp g\ppm g^{-1}\nonumber \\
&+&\frac{1}{6\pi} \epsilon^{ijk}\int\limits_{B,\partial B=\Sigma}
d^{3}y g^{-1}\partial_{i}gg^{-1}\partial_{j}gg^{-1}\partial_{k}g \nonumber \\
I_{A}&=&\frac{1}{2\pi}\intt [\Am^{R}g^{-1}\pp g-\Ap^{L}\ppm
gg^{-1}\nonumber \\
&+&\Ap^{L}g\Am^{R}g^{-1} \nonumber \\
&-&\frac{1}{2}(\Ap^{R}\Am^{R}+\Ap^{L}\Am^{L})].
\end{eqnarray}
We parametrize the group element $g \in SO(2,2)$ as in
\cite{bars}, i. e.,  $g=h\tau$, where $h \in SO(2,1)$ and $ \tau \in
SO(2,2)/SO(2,1)$.  The matrix $\tau$ has the form:

$$
\tau=\left(
\begin{array}{cc}
b&(b+1)\hat{X^{\nu}}\\
-(b+1)\hat{X_{\mu}}& \eta_{\mu}^{\nu}-(b+1)\hat{X_{\mu}}\hat{X^{\nu}}
\end{array}
\right),
$$
here $\hat{X^{\mu}}$ is the arrow $(x_{1},-x_{2},-x_{3})$,
$b=(1-x^{2})/(1+x^{2})$ and the indices are
contracted with the Minkowski metric $\eta_{\mu\nu}=diag(1,-1,-1)$. The
matrix $h$ is as follows:

$$
h=\left(
\begin{array}{cc}
1&0\\
0& h_{\mu}^{\nu}
\end{array}
\right),
$$
where $h_{\mu}^{\nu}=[(1+a_{\mu}^{\nu})(1-a_{\mu}^{\nu})^{-1}]$, with
$a_{\mu\nu}=-a_{\mu\nu}$. We consider the gauge transformation generated by

\begin{eqnarray}
T^{0}&=&\frac{1}{2}\left(
\begin{array}{cccc}
0&0&0&0 \\
0&0&1&0 \\
0&1&0&0 \\
0&0&0&0
\end{array}
\right), \nonumber \\
T^{1}&=&\frac{1}{2}\left(
\begin{array}{cccc}
0&0&0&0 \\
0&0&0&0 \\
0&0&0&1 \\
0&0&-1&0
\end{array}
\right).
\end{eqnarray}
We will impose gauge conditions on $h$, namely $a_{12}=a_{23}=0$. A
convenient substitution is $\cosh r=(1+a_{13}^{2})/(1-a_{13}^{2})$, which is
possible for $|a_{13}|<1$. The kinetic term of the WZW model can be represented
in a more compact form after the following change of variables which
diagonalyzes the part of the metric corresponding to the coset element $\tau$:

\begin{eqnarray}
x_{1}&=&t\cosh\te, \nonumber \\
x_{2}&=&t\sinh\te\cos\psi, \nonumber \\
x_{3}&=&t\sinh\te\sin\psi. \nonumber \\
\end{eqnarray}
In these new coordinates we obtain

\begin{eqnarray}
Tr\pd g\pu g^{-1}&=&\frac{8}{(\T)^{2}}\pd t\pu t \nonumber \\
&-&\frac{8t^{2}}{\T}\left(\pd \te\pu \te+\sh^{2}\te\pd\psi\pu\psi+\sin\psi\pd \te\pu r
\right. \nonumber \\
&+& \left. \frac{\sh 2\te}{2}\cos\psi\pd\psi\pu r\right)-2\pd r\pu r.
\end{eqnarray}

The Wess--Zumino term is by construction a closed three-form. According to
the Poincare lemma it can be rewritten locally as a two-form. For the case
under consideration we have:

\begin{equation}
\epsilon^{\mu\nu}b^{0}_{ij}\pd x_i \pdn x_{j}=\epsilon^{\mu\nu}
\frac{2t^{2}}{\T}(2\sin \psi \pd \te\pdn r+ \cos \psi\sinh 2\te\pd\psi\pdn r)
\end{equation}

The currents of the model, which are defined as $\Jm^{a}=Tr T^{a}
\ppm g g^{-1}$ and $\Jp^{a}=Tr T^{a} g^{-1}\pp g$   with $a=0,1$, are:

\begin{eqnarray}
\Jm^{0}&=&\frac{t^{2}}{\T}(4\cos\psi\cosh r\ppm \te \nonumber \\
&-&4\sinh \te(\sin\psi\cosh \te\cosh r-\sinh r\sinh \te)\ppm\psi) \nonumber \\
\Jm^{1}&=&\frac{t^{2}}{\T}(4\cos\psi\sinh r\ppm \te \nonumber \\
&-&4\sinh \te(\sin\psi\cosh \te\sinh r-\cosh r\sinh \te)\ppm\psi)  \nonumber \\
\Jp^{0}& =& \frac{t^{2}}{\T}(2\sinh^{2}\te\sin \psi \pp r \nonumber\\
&-&4\cos\psi\pp \te+2\sin\psi\sinh 2\te\pp\psi) \nonumber \\
\Jp^{1}&=& \frac{t^{2}}{\T}(2\sinh 2\te\cos\psi\pp r+4\sinh^{2}\te\pp\psi).
\end{eqnarray}

According to general  practice we now exclude gauge field in
order to read off the low energy content of the theory. The result is

\begin{equation}
I(g,A(g))=kI(g)-\frac{k}{2\pi}\innt \Jm^{a}(M^{ab})^ {-1}\Jp^{b},
\end{equation}
where $M^{ab}=Tr(T^{a}T^{b}-T^{a}gT^{b}g^{-1})$. The values of the inverse
matrix are
\begin{eqnarray}
(M^{-1})_{00}&=&\frac{2t^{2}}{\D(\T)}(\cosh 2\te\cosh r\nonumber \\
&+&\sin \psi\sinh 2\te\sinh r-1+\frac{2}{t^{2}}\sinh^{2}(r/2)) \nonumber \\
(M^{-1})_{01}&=&-\frac{2t^{2}}{\D(\T)}(\cosh 2\te(\sinh r+\sin\psi\cosh
r)+\frac{1}{t^{2}}\sinh r) \nonumber \\
(M^{-1})_{10}&=&-\frac{2t^{2}}{\D(\T)}\left(\sin\psi(\sinh 2\te\cosh r+\sin\psi\cosh
2\te\sinh r)+ \cos^{2}\psi \right.\nonumber \\
&-&\left.\frac{1}{t^{2}}\sinh r\right) \nonumber \\
(M^{-1})_{11}&=&\frac{2t^{2}}{\D(\T)}\left(\sin\psi(\sinh 2\te\sinh r+\sin\psi\cosh
2\te\cosh r)  \right. \nonumber \\
&+&\left.\cosh r\cos^{2}\psi+1-\frac{2}{t^{2}}\sinh^{2}(r/2)\right)
\end{eqnarray}
where $\D=det M^{ab}$.
The proper integration of the gauge field in the large $k$ (one-loop)
approximation includes shifting the dilaton field as follow:
\begin{eqnarray}
\hat{\Phi}&=&\hat{\Phi}_{0}+\ln det M^{ab}  \nonumber \\
&=&\hat{\Phi}_{0}                           \nonumber \\
&+&\ln\left(\frac{t^{2}}{\T}\cos^{2}\psi\sinh^{2}\te\cosh r
+\frac{2}{\T}\sinh^{2}(r/2) \right.         \nonumber \\
&+&\left.  \frac{t^{2}(t^{2}-1)}{(\T)^{2}}\cos^{2}\psi\sinh^{2}\te \right)
\end{eqnarray}

The whole low energy background of this theory consists as expected of a
metric and antisymmetric tensors as well as a dilaton field. In general
it does not have any evident symmetry. A very interesting feature of this
solution is that although for $t=0$ the metric has an initial singularity
the dilaton does not blow up at this point and equals
$\hat{\Phi}=2\ln\sinh (r/2)$. For small values of $t$ the metric is described
by eq. (4), which we rewrite as
\begin{eqnarray}
ds^{2}&=&dt^{2}-\frac{1}{4}dr^{2}-t^{2}(d\te^{2}+\sh^{2}\te d\psi^{2}) \nonumber \\
&-&t^{2}(\sin\psi d\te d r+\frac{\sh\te}{2}\cos\psi d\psi d r).
\end{eqnarray}
The first line describes a Kantowski-Sachs metric with negative curvature, the
second line is a modification.
This metric represents an expanding universe since the volume element
$$
\sqrt{-g}=\frac{1}{2}t^{2}\sinh\te
$$
constantly increases in the range of $t$ under consideration. The separation
between two observers is $t\Delta\te$ if only their $\te$ coordinates
differ
and $\Delta r$ if only their $r$ coordinates differ. Thus, the distances
measured only in the $\te$ and $\psi$ directions expand on a rate proportional
to $t$ while in the $r$ direction do not expand at all.
We can conclude that
the model describes an anisotropically expanding universe with the metric being
a modification of the Kantowski-Sachs metric with negative curvature.
It can also be checked that for large
values of $t$, i. e. $t>>1$, the metric tensor has the form:

\begin{equation}
ds^{2}=\frac{1}{t^{4}}dt^{2}-h_{ij}dx^{i}dx^{j},
\end{equation}
where $h_{ij}$ is independent of $t$ and must be read off from eqs. (4,6-8) in
the large $t$ limit, and $x^{i}=(r,\te,\psi)$.  After the substitution $T=1/t$
the metric becomes

\begin{equation}
ds^{2}=dT^{2}-h_{ij}dx^{i}dx^{j},
\end{equation}
which is evidently a static Universe.  We can conclude that the low energy
limit of the  analyzed conformal field theory is a universe that has
an initial singularity and then expands to a static limit. For intermediate
values of $t$ we are not able to give an exact cosmological
interpretation of the theory.  To consider this bosonic solution as part of a
heterotic string we have to follow the Kazama--Suzuki construction and add
left and right  fermions whose anomalies cancel among themselves. In the next
section we will construct generalized solutions of
this result.

\section{ Heterotic Coset Model techniques}
\subsection{ A Background with Electromagnetic fields}
In this subsection we present the simplest background that can be obtained
using heterotic coset model techniques. In fact, this is a cousin of the
monopole theory of Giddings Polchinski and Strominger \cite{gps} as presented
by Johnson \cite{john1}. We start with the right gauging of the WZW model,
which can be obtained from (1) taking $A^{L}=0$ and $A^{R}=A$:

\begin{equation}
I_{RGWZW}(g,A)=k I(g)+\frac{k}{2\pi}\int d^2 z Tr \Am g^{-1}\pp g
-\frac{k}{4\pi}\int d^2 z Tr \Ap\Am.
\end{equation}

Under the transformations

$$
g \to gh \quad \delta g=-gu \quad \delta A_a =-D_a u
=-\partial_a u-[A_a ,u],
$$
the action changes to
$$
\delta I(g,A)=\frac{k}{4\pi} \int d^2 z u(\pp \Am-\ppm \Ap).
$$
In the heterotic coset model this classical anomaly cancels
against the one-loop anomalies produced by the fermions. The anomaly of the
right fermions is fixed by supersymmetry since we require supersymmetry in
the right sector. The number of right fermions is $Dim(G)-Dim(H)=4$, they are
all minimally coupled to $\Am$ and produce an anomaly of the
form:

\begin{equation}
\delta I_{F}^{R}= \frac{1}{4\pi}\intt u F_{z\bar{z}}.
\end{equation}
The left moving fermions play the role of current algebra, are minimally
coupled to $\Ap$ with coupling constant $Q$ and produce an anomaly:

\begin{equation}
\delta I_{F}^{L}= -\frac{Q^{2}}{4\pi}\intt u F_{z\bar{z}}.
\end{equation}

The anomaly cancelation condition is

$$
k=Q^{2}-1
$$
Reaching the low energy limit is not as easy as in the previous section, the
problem is that now the level $k$ is related to the coupling $Q$. This
implies that the fermions affect the metric in the same order as the gauge
fields. To solve this puzzle and be able to integrate out the gauge field we
need a classically gauge invariant action.  This can be
achieve by bosonizing the fermions. The bosonization of the fermions is not
trivial.  Keeping in mind that taking
care of the anomalies is our main concern, it will be enough to find a
bosonic action with the same anomaly as the fermion system.  An anomalously
gauged WZW model at level $k=1$ exhibits the desired anomaly properties.
Nevertheless, following this direction we find that reading off the metric is
very complicated.  We can proceed as \cite{john1} and introduce a bosonic
action yielding the same anomaly,

\begin{equation}
I_{B}=\frac{1}{4\pi}\innt (\pp \phi^{a} -Q_{+}\Ap^{a})(\ppm \phi^{a}
-Q_{+}\Am^{a}) -Q_{-} \phi^{a} F_{z\bar{z}}^{a},
\end{equation}

here $Q_{\pm}=Q\pm 1$, $a=0,1$. This action under the transformations
$\delta\phi^{a}=\Qp u^{a}$ and $\delta A_{\pm}^{a}=\partial_{\pm}u^{a}$ yields
the desired anomaly. The index $a$ is contracted using the  metric
$\eta^{ab}=Tr T^{a}T^{b}=1/2 diag(1,-1)$. The total action that we get is:

\begin{eqnarray}
I^{total}&=&kI(g)+\frac{1}{4\pi}\innt\left(\pp\phi^{a}\ppm\phi^{a} \right.
\nonumber \\
&+&\Am^{a}(2k\Jp^{a}-(\Qp-\Qm)\pp\phi^{a}) \nonumber \\
&-&\Ap^{a}(\Qp+\Qm)\ppm\phi^{a}  \nonumber \\
&+&\left.\Ap^{a}\Am^{a}(-k+\Qp^{2})\right).
\end{eqnarray}
Since the gauge field is nondynamical and enters the action quadratically we
can integrate it obtaining

\begin{eqnarray}
I^{total}&=&kI(g)+\frac{1}{4\pi}\innt\left(\pp\phi^{a}\ppm\phi^{a} \right. \nonumber \\
&-&\left.\frac{(\Qp+\Qm)}{k-\Qp^{2}}\ppm\phi^{a}(2k\Jp^{a}-(\Qp-\Qm)\phi^{a}\right).
\end{eqnarray}

>From this action it is still not clear the way the metric should be affected by
the integration of the gauge field, to see this explicitly we have to
prepare the action  for refermionization or at least present it in a
suitable form.  We have to build a convenient term of the form
$\widetilde{D_{+}}\phi \widetilde{D_{-}} \phi $, forming this term uniquely
determines the metric and the gauge field.  We rewrite the total action as:

\begin{eqnarray}
I^{total}&=&kI(g)+\frac{\Qp^{2}-\Qm^{2}}{4\pi(k-\Qp^{2})}\innt
\pp\phi^{a}\ppm\phi^{a} \nonumber \\
&+&\frac{1}{4\pi}\innt\left((\pp
\phi^{a}-\frac{k(\Qp+\Qm)}{k-\Qp^{2}}\Jp^{a})( \ppm
\phi^{a}-\frac{k(\Qp+\Qm)}{k-\Qp^{2}}\Im^{a})\right. \nonumber \\
&-&\frac{k(\Qp+\Qm)}{k-\Qp^{2}}(\Jp^{a}\ppm \phi-\pp \phi\Im^{a}) \nonumber  \\
&-& \left. \frac{k^{2}(\Qp+\Qm)^{2}}{(k-\Qp^{2})^{2}}\Jp^{a}\Im^{a} \right).
\end{eqnarray}

where $\Im^{a}=Tr T^{a}g^{-1}\ppm g$. The last term is the one that affects
the part of the metric defined by $kI(g)$, the third term tells us that the
fermions will now be in the presence of an abelian field. The final
background is:

\begin{eqnarray}
ds^{2}&=&\frac{1}{(\T)^{2}}dt^{2}-\frac{t^{2}}{\T}h_{ij}dx^{i}dx^{j}
-\frac{t^{4}}{(\T)^{2}}\gamma_{ij}dx^{i}dx^{j}-2dr^{2} \nonumber \\
b_{\te r}&=&\frac{4t^{2}}{\T}\sin\psi \quad  b_{\psi r}=\frac{2t^{2}}{\T}\sinh 2\te\cos\psi   \nonumber \\
{\cal A}_{r}^{0}&=&\frac{2t^{2}}{\T}\sinh^{2}\te\sin 2\psi \quad  {\cal A}_{r}^{1}=\frac{2t^{2}}{\T}\sinh 2\te\cos\psi \nonumber \\
{\cal A}_{\te}^{0}&=&\frac{-4t^{2}}{\T}\cos\psi \quad  {\cal A}_{\te}^{1}=0 \nonumber \\
{\cal A}_{\psi}^{0}&=&\frac{2t^{2}}{\T}\sinh 2\te\sin\psi \quad  {\cal A}_{\psi}^{1}=\frac{4t^{2}}{\T}\sinh^{2}\te \nonumber \\
{\cal A}_{t}^{0}&=& 0 \qquad  {\cal A}_{t}^{1}=0 \nonumber \\
\hat{\Phi}&=&\hat{\Phi}_{0}+\ln(-k+\Qp^{2}).
\end{eqnarray}
As we see the dilaton field is trivially shifted by a constant, the expression
for the antisymmetric tensor is entirely determined by $I(g)$. As in the
previous section for small values of $t$, we have an anisotropically
expanding universe and for large values of $t$ we have a static universe. The
major difference with the vector gauge WZW of the previous section is the
presence of two sets of abelian gauge fields. Both electromagnetic fields
depend on time. As a function of time the electric and magnetic fields
are zero for $t=0$ and for large values of $t$ the electric field tends
to zero while the magnetic field goes to a constant. One feature of this
solution is that the value of the charge is fixed by the choice of $k$. In
the next subsection we will see that heterotic coset model techniques
provide an opportunity to avoid this restriction and construct
solutions whose charge is arbitrary in the sense that it is not constrained
by the choice of the level $k$.

\subsection{A background with arbitrary charge}
The next generalization of the presented construction consists of including a
parameter $\delta $ to move away for the restricting condition on $Q$. We now
start with an anomalously gauged WZW action eq. (1) for which
$t_{R}=T^{a}$ and $t_{L}=\delta T^{a}$ with $a=0,1$. The fermionic
action is the same as in the previous subsection and the anomaly cancellation
condition now reads as

\begin{equation}
k(1-\delta^{2})=Q^{2}-1
\end{equation}

Again we bosonize the fermionic sector to properly integrate out the gauge
field. The total action is now of the form:

\begin{eqnarray}
I^{total}&=&kI(g)+\frac{1}{4\pi}\innt \pp\phi^{a}\ppm\phi^{a} \nonumber \\
&+&\frac{1}{4\pi}\innt \left( \Am^{a}(2k\Jp^{a}-(\Qp-\Qm)\pp\phi^{a}) \right. \nonumber \\
&-&\Ap^{a}(2k\Jm^{a}+(\Qp+\Qm)\ppm\phi^{a}) \nonumber \\
&-& \left.
\Ap^{a}\Am^{b}(\eta^{ab}(k(1+\delta^{2})-\Qp^{2})-2k\delta
TrT^{a}gT^{b}g^{-1} ) \right).
\end{eqnarray}
Integrating out the gauge field and accommodating the action as in the
previous subsection we obtain:

\begin{eqnarray}
I^{total}&=&kI(g)+\frac{1}{4\pi}\innt(\pp\phi^{a}\ppm\phi^{b}\l^{ab} \nonumber \\
&-&4k^{2}\Jp^{a}\l^{ab}\Jm^{b}) \nonumber \\
&+&\frac{1}{4\pi}\innt\left((\pp\phi^{a}-2kQ\Jp^{b}\l^{ab}+2k\Ip^{b}\l^{ba})\right.  \nonumber \\
&\times &(\ppm\phi^{a}-2kQ\Im^{b}\l^{ab}+2k\Jm^{b}\l^{ba}) \nonumber \\
&-&k(\Qp+\Qm)\l^{ba}(\Jp^{a}\ppm\phi^{b}-\Im^{a}\pp\phi^{b}) \nonumber \\
&-&k(\Qp-\Qm)\l^{ba}(\Jm^{b}\pp\phi^{a}-\Ip^{b}\ppm\phi^{a}) \nonumber \\
&-&(k(\Qp+\Qm))^{2}\Jp^{i}\Im^{j}\l^{ai}\l^{aj}-(k(\Qp-\Qm))^{2}\Ip^{i}\Jm^{j}\l^{ia}\l^{ja}  \nonumber \\
&-&k^{2}\left.(\Qp^{2}-\Qm^{2})(\Jp^{i}\Jm^{j}\l^{ai}\l^{aj}+\Ip^{i}\Im^{j}\l^{ia}\l^{ja}) \right),
\end{eqnarray}
where $\Ip^{a}=Tr \pp g g^{-1}$ and $\l^{ab}$ is the inverse matrix to
$M^{ab}=\eta^{ab}(k(1+\delta^{2})-\Qp^{2})-2k\delta Tr T^{a}gT^{b}g^{-1}$.
The dilaton is shifted by

$$
\hat{\Phi}=\hat{\Phi}_{0}+\ln det M^{ab}.
$$
As opposed to the previous subsection the dilaton shift is not trivial.
Reading off the background is essentially the same as in the previous
subsection.  The general result is very complicated and so we outline the
relevant characteristics of this solution. For $\delta=0$ the background is
reduced to the one of the previous subsection and for
$\delta=1$ we recover the results of the previous section for the anomaly
free gauged WZW model. From here we can already conjecture that this is
another cosmological theory for some values of $\delta$ and in the limits of
large and small $t$.  For arbitrary values of $\delta$ one can check that the
metric indeed behaves the same as the other presented backgrounds. The
two electromagnetic fields present the same asymptotical behavior in time as
in the previous subsection, but do not depend on the time coordinate
as simply as in the previous case.

\section{Conclusions}
We have constructed some intrinsically four--dimensional heterotic string
solutions. The low energy limit of the first solution (vector gauged
WZW model) includes metric and antisymmetric tensors and dilaton field, this
solution describes an anisotrpically expanding universe for small values of
$t$ and a static universe for large values of $t$; it also has an initial
singularity but it is remarkable that the dilaton does not blow up at this
singularity.  The other solutions, apart from these characteristics  possess
two sets of electromagnetic fields, in one case with a fixed value of the
charge and in the other with arbitrary charge.  We conjecture that the charged
solutions are possibly extremal solutions since we actually have very few
parameters; in the first case all the parameters are fixed, in the second we
only have one free parameter, the charge $Q$ (or $\delta$).  An interesting
unanswered question is to what extend these solutions can be obtained as low
energy limits of higher dimensional theories. Given that for large values of
$t$ the metric describes a static universe and has one timelike killing
vector it seems interesting to relate the properties of these solutions to
the symmetries of the low energy string effective action \cite{gal} and
determine which of them can be obtained starting from other simpler known
solutions.
\begin{center}
\large Acknowledgments
\end{center}
I am very grateful to my advisor D. V. Galtsov for suggesting me this theme
and guidance throughout the work on this paper. I have also benefited from
discussions with M. I. Iofa on related problems.

\end{document}